# Growth of Shock-Induced Solitary Waves in Granular Crystals

M. Arif Hasan, Sia Nemat-Nasser

Center of Excellence for Advanced Materials

Department of Mechanical and Aerospace Engineering

Jacobs School of Engineering

University of California, San Diego

9500 Gilman Drive, La Jolla, California, 92093, USA

mdhasan@ucsd.edu, sia@ucsd.edu

## ABSTRACT

Solitary waves (SWs) are generated in monoatomic (homogeneous) lightly contacting spherical granules by an applied input force of any time-variation and intensity. We consider finite duration shock loads and focus on the transition regime that leads to the formation of SWs. Based on geometrical and material properties of the granules and the properties of the input shock, we provide explicit analytic expressions to calculate the peak value of the compressive contact force at each contact point in the transition regime that precedes the formation of a primary solitary wave. We also provide explicit expressions to estimate the number of granules involved in the transition regime and show its dependence on the characteristics of the input shock and material/geometrical properties of the interacting granules. Finally, we assess the accuracy of our theoretical results by comparing them with those obtained through numerical integration of the equations of motion.

1. **Introduction**

Granular crystals are ordered aggregates of contacting granules with highly nonlinear elastic interaction and tunable wave propagation properties [1,2]. When neighboring granules are compressed, the interactions are nonlinear, non-linearizable Hertzian, whereas in the absence of precompression, the granules may separate and provide additional source of nonlinearity due to subsequent collisions. Analytical, numerical, and experimental approaches have been used to understand such complex dynamic responses of granular crystals [1,3–7]. In the absence of an external prestress, the system is strongly nonlinear and acts as a "sonic vacua", lacking the characteristics of linear sound speed [1]. They can also be designed as phononic crystals [2,8]. Due to such unique properties, ordered arrays of granular crystals have become an intensely popular research area [9,10] and have found many engineering applications [11,12]. The media also supports solitary waves [1,4,5,10,13,14], discrete breathers [15–17,17], and display acoustic pass- and stop- bands [7,18].

The idea of employing granular crystals as passive energy absorbers for shock mitigation or blast protection is traced back of the pioneering work by Nesterenko [1] in 1983. Later, diatomic chain, tapered chain (either forward or backward linearly tapered or exponentially tapered) are also studied by several researchers both theoretically (mostly by using binary collision approximations), numerically, and experimentally [10,19] as a shock protector. The main principal of shock wave mitigation employing granular media is to use the material deformation to absorb energy. Most of these studies, however, consider either impulsive loading [6] or harmonic excitation [7] as an input to the system. In reality a shock wave has a sharp rise followed by a slow decay that cannot be approximated by an impulse or a harmonic input force.

Blast loading can arise from contact or air explosion [20]. The duration of a shock loading can range from several microseconds to milliseconds [20]. It is also reported that low explosives (e.g. propellants and gun powders) generate pressure pulses which are usually of longer duration in comparison to the high explosives [21]. For a broad class of finite duration input shocks of sharp rise and slow decay, we have shown how *a priori* one can establish quantitative relations between the input force and the peak value, the linear momentum, and other related properties of the resulting solitary waves [22]. Furthermore, in [23] we have provided explicit analytical expressions to calculate the time-variation of the displacement, the velocity, the acceleration of individual grains, and the compressive contact force between neighboring granules, as well as the linear momentum, the total energy, the equivalent (or effective) mass, and the effective velocity of the resulting SW. In the current paper, we give explicit expressions to estimate the number of granules and the value of the contact forces in the transition regime for input shocks of finite rise and decay durations, and assess the accuracy of these estimates through numerical simulations. As is known [1,24], under an impulsive load, a SW forms in a monoatomic granular crystal at approximately 7 or 10 granules diameter from the loading end. In general, however, both the number of granules and the magnitude of the contact forces will depend on the characteristics of the applied force and the properties of the granules, as is discussed later in detail.

## 2. System under Consideration

We seek to explore the response of a system of lightly contacting identical spherical elastic granules under shocks with finite rise and decay times. The governing equations of motion are,

$$m\ddot{u}_1 = F(t;t_0) - \left[\frac{\sqrt{2RE}}{3(1-v^2)}\right](u_1 - u_2)_+^{3/2}; \quad m\ddot{u}_N = \left[\frac{\sqrt{2RE}}{3(1-v^2)}\right](u_{N-1} - u_N)_+^{3/2},$$

$$m\ddot{u}_i = \left[\frac{\sqrt{2RE}}{3(1-v^2)}\right]\left[(u_{i-1} - u_i)_+^{3/2} - (u_i - u_{i+1})_+^{3/2}\right]; \quad i = 2,\cdots,N-1, \quad (1)$$

where $u_i$ is the displacement of granule $i$, $N$ is the total granular, $m$ is the mass, $E$ is the Young's modulus, $v$ is the Poisson's ratio, and $R$ is the radius of a typical granule, respectively. The subscript (+) indicates that the granules are cohesion free (do not support tension [25]). Here, $F(t;t_0)$ is the applied shock given by,

$$F(t;t_0) = F_0[F_R(t)H(t_0 - t) + F_D(t)H(t - t_0)H(nt_0 - t)]; \quad t \in (0,\infty], \quad (2)$$

where $t_0$ is the rise time, $F_0$ is the intensity of the input shock, $nt_0$ is the total pulse duration, and $F_R(t)$ and $F_D(t)$ represent shock profiles during rise and decay periods, respectively. As in [22], we consider *shock inputs* with steep rise and slow decay durations. For example, when an striker (e.g., a relatively large granule) impacts a chain of contacting granules, the resulting input compressive force would have a steep rise profile followed by a long relaxation tail [26]. Hence, the characteristic time of unloading is generally greater than the rise time. This type of steep rise and slow decay shock pulses are known as *monopolar pulse* [27]. Here, we focus our attention on input forces with rise and decay profiles, $F_R(t)$ and $F_D(t)$, that satisfy [22],

$$\int_{t_0}^{nt_0} F_D(t)dt \geq \int_0^{t_0} F_R(t)dt; \quad \int_0^{t_0} F_R(t)dt \geq t_0/2; \quad \int_{t_0}^{nt_0} F_D(t)dt \leq (n-1)t_0/2.$$

## 3. Problem Statement

Upon the application of a finite duration shock load, the peak value, $F_{i_{max}} = max[F_i(t)]$, of the intergranular Hertzian [28] contact force,

$$F_i(t) = \left[\frac{\sqrt{2RE}}{3(1-\nu^2)}\right][(u_{i-1} - u_i)^{3/2}]; \quad i = 2, \cdots, N-1,$$

decreases, remain constant, or increases over an initial set of, say, $p$ granules until eventually a SW is formed at steady state regime [22] where

$$F_{i_{max}} \simeq F_{SW_{max}}; \quad \text{for } i \geq p.$$

The maximum amplitude, $F_{SW_{max}}$, can be predicted *a priori* by applying the analytical techniques developed in [22]. In the current paper, we consider a broad class of finite duration input shocks of sharp rise and slow decay which produces a primary SW such that $F_{SW_{max}}$ can be either greater or less than $F_0$, and seek to estimate: (i) the number, $p$, of the granules in the transition regime, and (ii) the peak amplitude, $F_{i_{max}}$, of the compressive contact force at each pair of contacting granules within the transition regime.

## 4. Analytical Results

### 4.1. Number, $p$, of the granules in the transition regime

By using the rise time, $t_0$, as the unit of time, the input linear momentum can be calculated by integrating the input force as follows:

$$L_{Applied}(t_0) = \left[\int_0^{t_0} F_R(t)dt + \int_{t_0}^{nt_0} F_D(t)dt\right]F_0.$$

In [23] we have shown that the linear momentum carried by a single SW of peak amplitude $F_{SW_{max}}$ is given by,

$$L_{SW} \simeq \frac{139}{32}\left[\frac{(1-\nu^2)R^4\rho^{\frac{3}{2}}}{E}\right]^{\frac{1}{3}} F_{SW_{max}}^{\frac{5}{6}}.$$

Keeping $F_0$ and $n$ constant, we find a new unit of time, $\hat{t}_0$, such that [22]

$$L_{Applied}(\hat{t}_0) = \left[\int_0^{\hat{t}_0} F_R(t)dt + \int_{\hat{t}_0}^{n\hat{t}_0} F_D(t)dt\right]F_0 \simeq \frac{139}{32}\left[\frac{(1-\nu^2)R^4\rho^{\frac{3}{2}}}{E}\right]^{\frac{1}{3}} F_0^{\frac{5}{6}}.$$

Now, we define the ratio of the applied linear momentum to the linear momentum corresponds to $\hat{t}_0$, by

$$LM_{ratio} = L_{Applied}(t_0)/L_{Applied}(\hat{t}_0).$$

By its definition, when $LM_{ratio} \simeq 1$, or when $t_0 \simeq \hat{t}_0$ then the entire input linear momentum would be used to form a SW of the peak amplitude equal to that of the input force, $F_0$, and hence a SW forms immediately. As is reported in the literature, the width of such SWs is about $(\sqrt{10}\pi)R$ i.e. approximately 5 granules diameter [1,29]. We have verified this using our numerical approach, as is discussed below in Section 5.1.

### 4.1.1. Estimation of $p$ when $LM_{ratio} > 1$

For a given input force profile with a total pulse duration $nt_0$, if $nt_0 \simeq n\hat{t}_0$, then the resulting SW spans over 5 granules. But if $nt_0 > n\hat{t}_0$, then additional granules corresponding to the extra linear momentum associated with the pulse duration $(nt_0 - n\hat{t}_0)$ are required before a full SW is formed. Assuming that after the time period $n\hat{t}_0$, the speed of the wave is approximately equal to the speed of a fully developed SW, the number of granules within the transition range, $p$, can be estimated,

$$p \simeq 5 + \frac{n(t_0 - \hat{t}_0)V_{SW}}{2R}, \qquad for \quad LM_{ratio} > 1, \tag{3a}$$

where $V_{SW}$ is the speed of SW, given by (see equation (2) of [23]),

$$V_{SW} \simeq \frac{7}{10}\left[\frac{E}{R(1-\nu^2)\rho^{3/2}}\right]^{1/3}(F_{SW_{max}})^{1/6}.$$

### 4.1.2. Estimation of $p$ when $LM_{ratio} < 1$

For a given input force profile with a total pulse duration $nt_0$, if $nt_0 < n\hat{t}_0$, then the first granule (and possibly other granules close to the application point of the input force) would be ejected, imparting additional momentum to the rest of the granules [22]. The expression for this additional momentum is (see equation (5h) of Reference [22]),

$$\Delta L \simeq 2 \int_{n\hat{t}_0-(\hat{t}_0-t_0)}^{n\hat{t}_0} F(t;\hat{t}_0).$$

Again by assuming that after the time period $n\hat{t}_0$, the speed of the wave is approximately equal to the speed of a fully developed SW, the number of granules within the transition range, $p$, can be estimated by,

$$p \simeq 5 + \frac{[n\hat{t}_0 - \{n\hat{t}_0 - (\hat{t}_0 - t_0)\}]V_{SW}}{2R} \simeq 5 + \frac{(\hat{t}_0 - t_0)V_{SW}}{2R}, \quad for \quad LM_{ratio} < 1. \quad (3b)$$

It is clear from equations (3a,b) that, for a finite duration input shock, the number of granules require to form a steady-state SW depends on the material and geometrical properties of the contacting granules as well as on the input shock force.

### 4.2. Peak contact forces, $F_{i_{max}}$

In dimensionless parameters, we express the time variation of the contact force between the $i^{th}$ and $(i-1)^{th}$ granule as follows:

$$\bar{F}_i(\tau) = [\bar{u}_{i-1}(\tau) - \bar{u}_i(\tau)]^{3/2} \simeq \bar{F}_{i_{max}}[S(\tau/\bar{T})]^{3/2}, \quad (4a)$$

where $\bar{u}_i = u_i/R$, $\tau = t/\beta$, and $\beta = \left[2\sqrt{2}\pi\rho R^2(1-\nu^2)/E\right]^{1/2}$. $S(\tau/\bar{T}) = [1 + 0.439(\tau/\bar{T})^2 + 0.0994(\tau/\bar{T})^4 + 0.0147(\tau/\bar{T})^6 + 0.0013(\tau/\bar{T})^8]^{-2}$ is the Padé approximation of the temporal variation of the SW contact force [22,23], with $max\left|[S(\tau/\bar{T})]^{3/2}\right| = 1$ and $\bar{T}$ is the normalized peak-to-peak temporal shift of the SW contact force. Before a SW is formed, both the temporal profile and the time shift depend on the momentum of the applied force, and, as can be surmised from the results given in [23], we have,

$$\bar{T}_i \simeq \left(1.077 \bar{F}_{i_{max}}^{-\frac{1}{6}}\right)(LM_{ratio})^{\frac{25(p-i)}{16\sqrt{np}}}, \tag{4b}$$

for $2 \leq i \leq p$. Here, $\bar{T}_1 = \bar{t}_0$, $\bar{T}_p \simeq \bar{T}$, and $F_{1_{max}} = F_0$. We have verified expression (4b) through extensive numerical simulations.

### 4.2.1. Estimation of $F_{i_{max}}$ when $LM_{ratio} > 1$

According to Reference [22], if $LM_{ratio} > 1$, then a percentage of the input momentum, $L_{Applied}$, is used to generate the first (primary) SW with a peak amplitude $\bar{F}_{SW_{max}}$. Considering only the primary SW, and noting from [23] that the total linear momentum of a primary SW is, $\frac{53}{32}\bar{F}_{SW_{max}}^{5/6}$, we write,

$$\int_0^\infty \bar{F}_i(\tau)d\tau \simeq \int_0^\infty \bar{F}_{SW}(\tau)d\tau \simeq \int_{-\infty}^\infty \bar{F}_{i_{max}}\left[S\left(\frac{\tau}{\bar{T}_i}\right)\right]^{\frac{3}{2}} d\tau \simeq 1.543 \bar{F}_{i_{max}} \bar{T}_i \simeq \frac{53}{32}\bar{F}_{SW_{max}}^{\frac{5}{6}}. \tag{5}$$

Upon substitution for $\bar{T}_i$, we finally obtain,

$$F_{i_{max}} \simeq \left[(LM_{ratio})^{\frac{15(i-p)}{8\sqrt{np}}}\right] F_{SW_{max}} \tag{6a}$$

for $2 \leq i \leq p$ and $F_{1_{max}} = F_0$.

### 4.2.2. Estimation of $F_{i_{max}}$ when $LM_{ratio} < 1$

As mentioned above in section 4.1.2. and also in Reference [22], if $LM_{ratio} < 1$, the first granule (and possibly other granules close to the application point of the input force) would be ejected, imparting additional momentum to the rest of the granules, which the calculation of $F_{i_{max}}$ based on equation (6a) as outlined above, does not include. Therefore, the total input momentum to form $F_{i_{max}}$ would be $L_{Applied}(t_0) + \Delta L$ from which we obtain,

$$F_{i_{max}} \simeq \left[\left\{\frac{L_{Applied}(t_0) + \Delta L}{L_{Applied}(\hat{t}_0)}\right\}^{\frac{15(i-p)}{8\sqrt{np}}}\right]F_{SW_{max}} \simeq \left[\left\{LM_{ratio} + \frac{\Delta L}{L_{Applied}(\hat{t}_0)}\right\}^{\frac{15(i-p)}{8\sqrt{np}}}\right]F_{SW_{max}}. \quad (6b)$$

## 5. Numerical Validation

We use the three profiles shown in Figure 1, and defined in Table 1, for the input force and vary the linear momentum of each profile using various values for the momentum ratio, $LM_{ratio}$, to asses the accuracy of the analytical estimate of the number of granules and the peak contact forces in the transition regime.

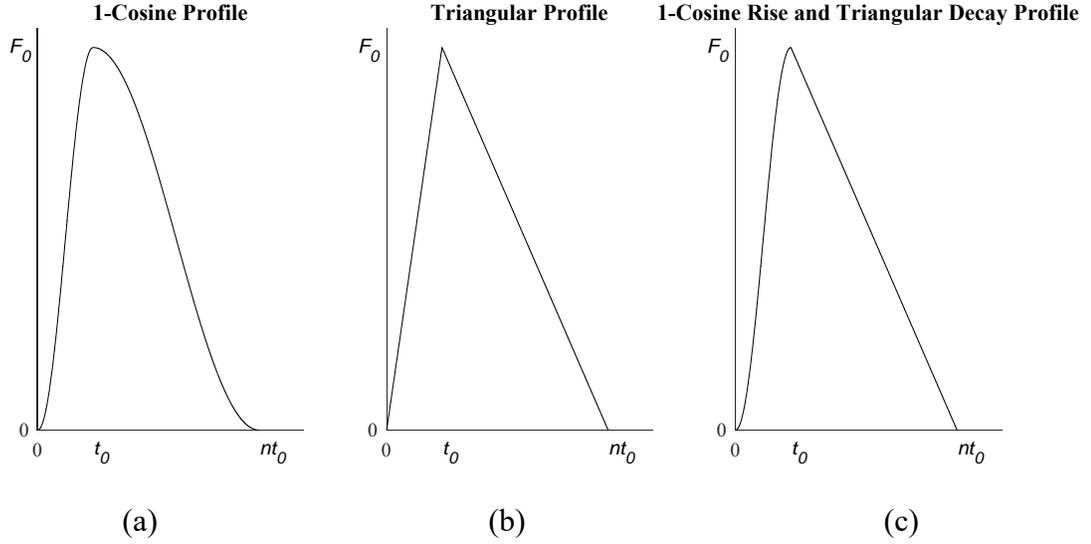

(a)      (b)      (c)

Figure 1. Various input shock types: (a) "1-Cosine shock profile" with $F_R(t;t_0) = \{1 - cos(\pi t/t_0)\}/2$ and $F_D(t;t_0) = \{1 - cos[\pi[t + (n-2)t_0]/[(n-1)t_0]]\}/2$, (b) "Triangular shock profile" with $F_R(t;t_0) = t/t_0$ and $F_D(t;t_0) = (nt_0 - t)/[(n-1)t_0]$, and (c) "1-Cosine Rise and Triangular Decay shock profile" with $F_R(t;t_0) = \{1 - cos(\pi t/t_0)\}/2$ and $F_D(t;t_0) = (nt_0 - t)/[(n-1)t_0]$.

Table 1. Input shock types.

| Shock Type | $F_R(t; t_0)$ | $F_D(t; t_0)$ |
|---|---|---|
| 1-Cosine | $\frac{1}{2}\left\{1 - \cos\left(\pi\frac{t}{t_0}\right)\right\}$ | $\frac{1}{2}\left\{1 - \cos\left[\pi\frac{t + (n-2)t_0}{(n-1)t_0}\right]\right\}$ |
| Triangular | $\frac{t}{t_0}$ | $\frac{(nt_0 - t)}{(n-1)t_0}$ |
| 1-Cosine rise and Triangular decay | $\frac{1}{2}\left\{1 - \cos\left(\pi\frac{t}{t_0}\right)\right\}$ | $\frac{(nt_0 - t)}{(n-1)t_0}$ |

First we numerically verify the width of a SW and show that it is approximately 5 granules diameter, as reported in [1,29]. Then we compare the analytical estimate of the number of granules, *p*, within the transition regime with the numerical result, using the input shock types defined in Table 1 and shown in Figure 1. Finally, we compare the peak contact forces given by equations (6a,b) with the corresponding results obtained numerically.

## 5.1. Width of a SW

As is reported in Ref. [23], once a SW is formed, the velocity profile of each granule in dimensionless unit is given by,

$$\bar{V}_{granule}(x) \simeq \bar{F}_{SW_{max}} \int^{x} \left\{[S(\tilde{x})]^{\frac{3}{2}} - [S(\tilde{x} - 1)]^{\frac{3}{2}}\right\} d\tilde{x},$$

where $\bar{V}_{granule}(x)$ is the spatial variation of the velocity of the granules, $\bar{F}_{SW_{max}}$ is the normalized peak amplitude of the compressive contact force, and $S(x)$ is the profile of the normalized compressive contact force, given by, $S(x) = (1 + 0.439x^2 + 0.0994x^4 + 0.0147x^6 + 0.0013x^8)^{-2}$.

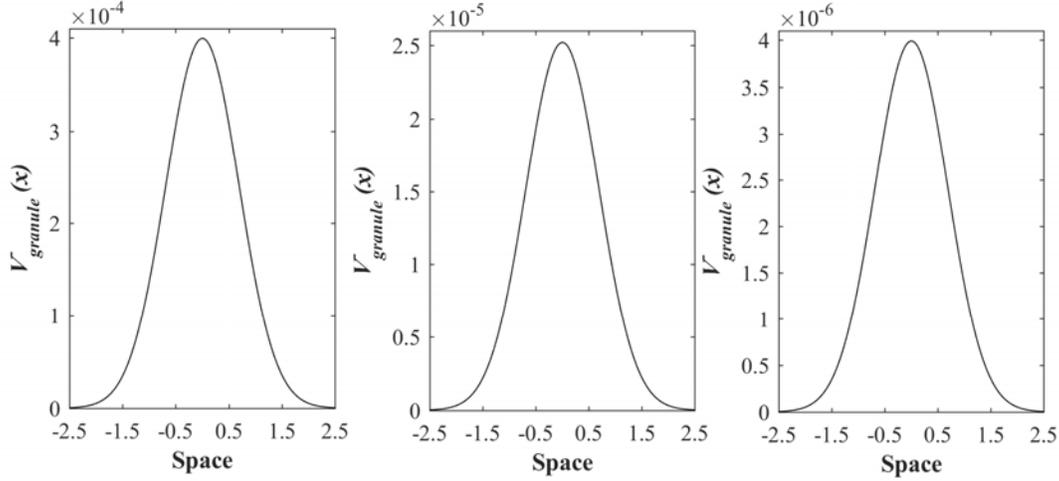

Figure 2. Granule's velocity profile in space; note the independence of the width of SW on the height; (a) large peak velocity amplitude, (b) intermediate peak velocity amplitude, and (c) small peak velocity amplitude.

In Figure 2, we have plotted granule's velocity profile in space for three different peak velocities; (a) large peak velocity, (b) intermediate peak velocity, and (c) small peak velocity. Figure 2 clearly shows that the width of a SW is independent of its peak value, and spans approximately 5 granules.

## 5.2. Number of granules, $p$, and peak contact forces, $F_{i_{max}}$, in the transition regime

In the numerical calculation, a force with the peak value $F_0 = F_{1_{max}} = 100N$ is applied to the first granule of the array and equations (1) are numerically integrated using vectorized fourth-order Runge-Kutta time integration scheme with the following parameters: $E = 193 GPa, \rho = 7958 kg/m^3, \nu = 0.3, R = 4.75mm$ [standard material properties]. For each case, we vary the applied linear momentum $L_{Applied}(t_0)$ of the input force by varying the given rise time $(t_0)$, and keeping $F_0$ and $nt_0$ fixed. The resulting responses are shown in Figure 3.

In Figure 3a, peak compressive contact forces, $F_{i_{max}}$, and in Figure 3b the total number of granules, $p$, are plotted for different linear momentum ratio, $LM_{ratio}$. It is clear from the figure that, after an initial transition regime, the peak amplitude of the contact force, $F_{i_{max}}$, reaches a steady state constant value, $F_{SW_{max}}$. The magnitude of $F_{SW_{max}}$ is dependent on the linear momentum ratio, $LM_{ratio}$, and, in the present cases, can be either higher or lower than the peak value of the input force $F_0$. From Figure 3b we see that greater $LM_{ratio}$ requires larger number of granules in the transition regime before a SW is formed.

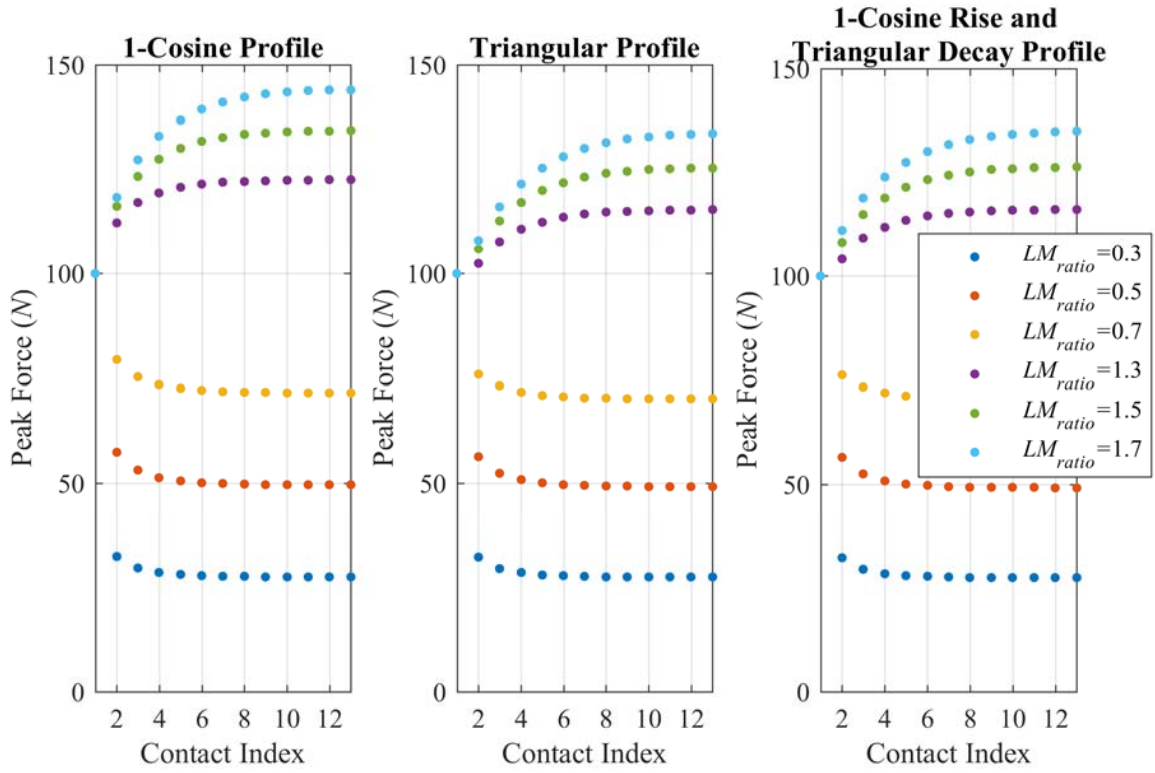

(a)

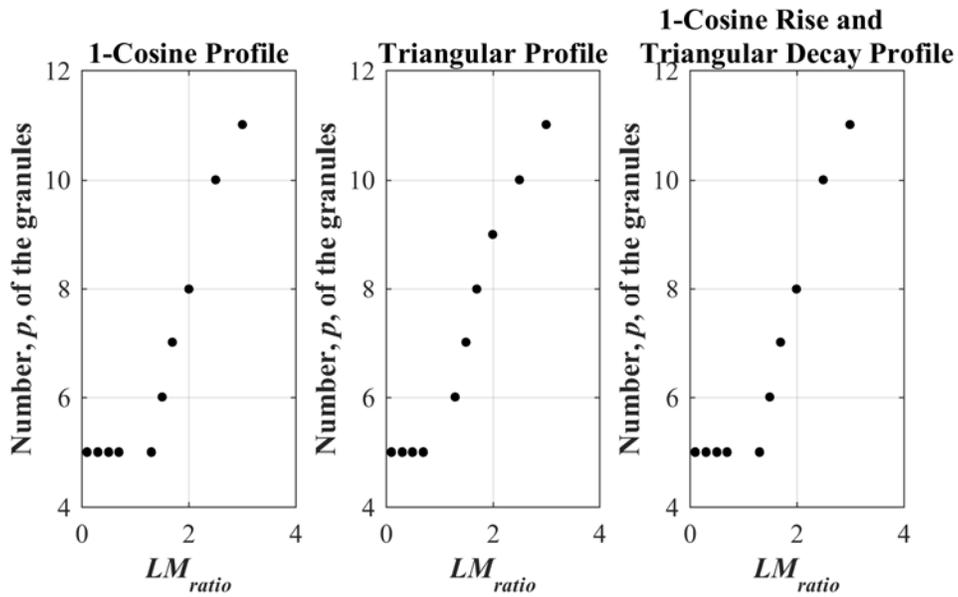

(b)

Figure 3. (a) Peak value of compressive contact forces, $F_{i_{max}}$, exerted at each neighboring granules, and (b) Total number of granules, $p$, to form SWs, for different linear momentum ratio, $LM_{ratio}$, and different input types. System parameters: $F_0 = F_{1_{max}} = 100N, \ nt_0 = 4t_0$.

In Table 2, the numerically obtained values of $p$ are compared with analytical ones for different linear momentum ratio, $LM_{ratio}$. As is seen, expressions (3a,b) provides reasonably accurate values of $p$. Further, Tables 3 and 4 compares the values of $F_{i_{max}}$ given analytically by equations (6a,b) with the corresponding values obtained by the numerical solution of equations (1). As is seen errors are quite small.

Table 2. Comparison between exact and approximate values of number, $p$, of the granules in the transition regime. System parameters: $F_0 = F_{1_{max}} = 100N, \ E = 193GPa, \rho = 7958kg/m^3, \nu = 0.3, \ R = 4.75mm$.

| Shock Type | $(LM_{ratio}, n)$ | Number, $p$, of the granules in the transition regime | |
|---|---|---|---|
| | | Approximate (Analytic-Equations (3a,b)) | Exact (Numerical-Equation (1)) |
| 1-Cosine | (0.6,6) | 5.2 | 5 |
| | (2.5,8) | 10.0 | 9 |
| Triangular | (0.4,10) | 5.2 | 5 |
| | (1.8,6) | 7.6 | 8 |
| 1-Cosine rise and Triangular decay | (0.8,8) | 5.1 | 5 |
| | (2.0,2) | 8.3 | 9 |

Table 3. Comparison between exact and approximate values of $F_{i_{max}}$, when $F_{i_{max}} < F_0$. System parameters: $F_0 = F_{1_{max}} = 100N, E = 193GPa, \rho = 7958kg/m^3, \nu = 0.3, \ R = 4.75mm$.

| Shock Type | $(LM_{ratio}, n)$ | Percentage of error in $F_{i_{max}}$ using equation (6a) | | | |
|---|---|---|---|---|---|
| | | $F_{2_{max}}$ | $F_{3_{max}}$ | $F_{4_{max}}$ | $F_{5_{max}}$ |
| 1-Cosine | (0.7,8) | 1.8 | 1 | 0.7 | 0.6 |
| | (0.5,6) | 2.3 | 1.3 | 0.2 | 0.8 |
| | (0.3,2) | 3.6 | 1.3 | 0.8 | 0.9 |
| Triangular | (0.6,2) | 0.6 | 0 | 0.3 | 0.7 |
| | (0.7,4) | 2.1 | 0.6 | 0 | 0.6 |
| | (0.4,8) | 3.5 | 1.9 | 0.4 | 0.9 |
| 1-Cosine rise and Triangular decay | (0.5,8) | 1 | 0.6 | 0 | 0.8 |
| | (0.4,2) | 3.8 | 1.3 | 0.8 | 0.9 |
| | (0.7,4) | 1.9 | 0.5 | 0 | 0.6 |

Table 4. Comparison between exact and approximate values of $F_{i_{max}}$, when $F_{i_{max}} > F_0$. System parameters: $F_0 = F_{1_{max}} = 100N, E = 193GPa, \rho = 7958kg/m^3, \nu = 0.3, R = 4.75mm$.

| Shock Type | $(LM_{ratio}, n)$ | Percentage of error in $F_{i_{max}}$ using equation (6b) | | | | | | | |
|---|---|---|---|---|---|---|---|---|---|
| | | $F_{2max}$ | $F_{3max}$ | $F_{4max}$ | $F_{5max}$ | $F_{6max}$ | $F_{7max}$ | $F_{8max}$ | $F_{9max}$ |
| 1-Cosine | (1.4,2) | 0.7 | 0.5 | 0.6 | 0.7 | 0.8 | | | |
| | (1.6,4) | 0.5 | 1 | 0.9 | 0.7 | 0.6 | 0.6 | | |
| | (1.7,2) | 0.2 | 2.8 | 2.4 | 1.6 | 1 | 0.8 | 0.8 | |
| Triangular | (1.4,4) | 2.8 | 2.3 | 1.6 | 1.1 | 0.9 | | | |
| | (1.6,6) | 3.9 | 3.3 | 2.5 | 1.7 | 1.1 | 0.8 | | |
| | (2,4) | 1 | 3.6 | 3.3 | 2.5 | 1.7 | 1.2 | 0.9 | 0.7 |
| 1-Cosine rise and Triangular decay | (1.4,6) | 3.8 | 2.5 | 1.7 | 1.1 | 0.8 | | | |
| | (1.7,4) | 0.3 | 1.7 | 1.7 | 1.3 | 1 | 0.9 | | |
| | (1.9,2) | 1.1 | 3.7 | 3.6 | 2.6 | 1.7 | 1.2 | 0.9 | 0.8 |

## 5.3. Alternative estimation of peak contact forces in transition regime

We normalize the position of each granule, which is defined by the index $i$ and is measured from the first granule, as well as the associated peak contact force, $F_{i_{max}}$, as follows:

$$j = \frac{i-2}{p-1}; \quad i = 2, \cdots, p+1$$

$$\bar{F}_{i_{max}} = 1 - \frac{F_{i_{max}} - F_{2_{max}}}{F_{SW_{max}} - F_{2_{max}}}; \quad i = 2, \cdots, p, \qquad (7a, b)$$

where $1 \leq j \leq 0$ and $1 \leq \bar{F}_{i_{max}} \leq 0$ are the normalized contact index and normalized peak contact forces, respectively. In Figure 4 we have plotted the normalized contact forces as function of the normalized contact index. Remarkably, when $LM_{ratio} > 1$, the results nicely fit the following 3$^{rd}$ degree polynomial (Figure 4b):

$$\bar{F}_{i_{max}} = 1 - 2.5j + 2.4j^2 - 0.9j^3. \qquad (8)$$

For the case of $LM_{ratio} < 1$ (Figure 4a), since the first granule (and possibly other granules close to the application point of the input force) ejects, hence the above polynomial does not fit well, nonetheless the errors are insignificant as can be seen in the following section 5.3.1.

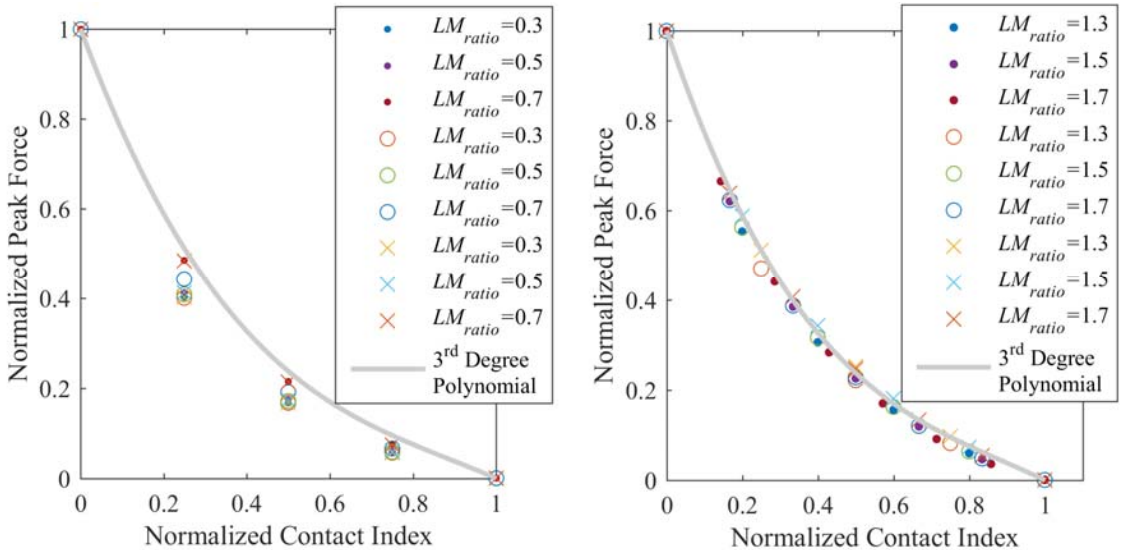

(a) $LM_{ratio} < 1$          (b) $LM_{ratio} > 1$

Figure 4. Normalized peak contact force, $\bar{F}_{i_{max}}$, against normalized contact index, $j$, for different linear momentum ratio, $LM_{ratio}$, and different input force profiles. Here, the exact numerical values for "1-Cosine shock profile" are denoted by "·", those for "Triangular shock profile" are denoted by "*o*", and those for "1-Cosine Rise and Triangular Decay shock profile" are denoted by "*x*". System parameters: $F_0 = F_{1_{max}} = 100N$, $nt_0 = 4t_0$; (a) $LM_{ratio} < 1$, and (b) $LM_{ratio} > 1$.

Combining equations (7a,b) and (8), we obtain,

$$F_{i_{max}} = F_{2_{max}} + \frac{(F_{SW_{max}} - F_{2_{max}})}{10}\left[25\left(\frac{i-2}{p-1}\right) - 24\left(\frac{i-2}{p-1}\right)^2 + 9\left(\frac{i-2}{p-1}\right)^3\right], \quad (9)$$

where $i = 2, \cdots, p+1$, and $F_{1_{max}} = F_0$. To use this equation, we need the value of $F_{2_{max}}$, which we can estimate using equations (6a,b).

### 5.4. Numerical validation of equations (6) and (9)

To estimate the peak contact forces in the transition regime from equations (6) and (9), we only need to calculate the total number of granules, *p*, in the transition regime. In Figure 5, we compare the peak contact forces in the transition regime obtained numerically with those obtained analytically, where "·" denotes the exact numerical value, "o" represents the result given by Equation (6), and "*" represents the result obtained from Equation (9). It is clear from Figure 5 that both Equations (6) and (9) can be successfully used to estimate the peak contact forces in the transition regime for an array of contacting granules under finite duration shock loads.

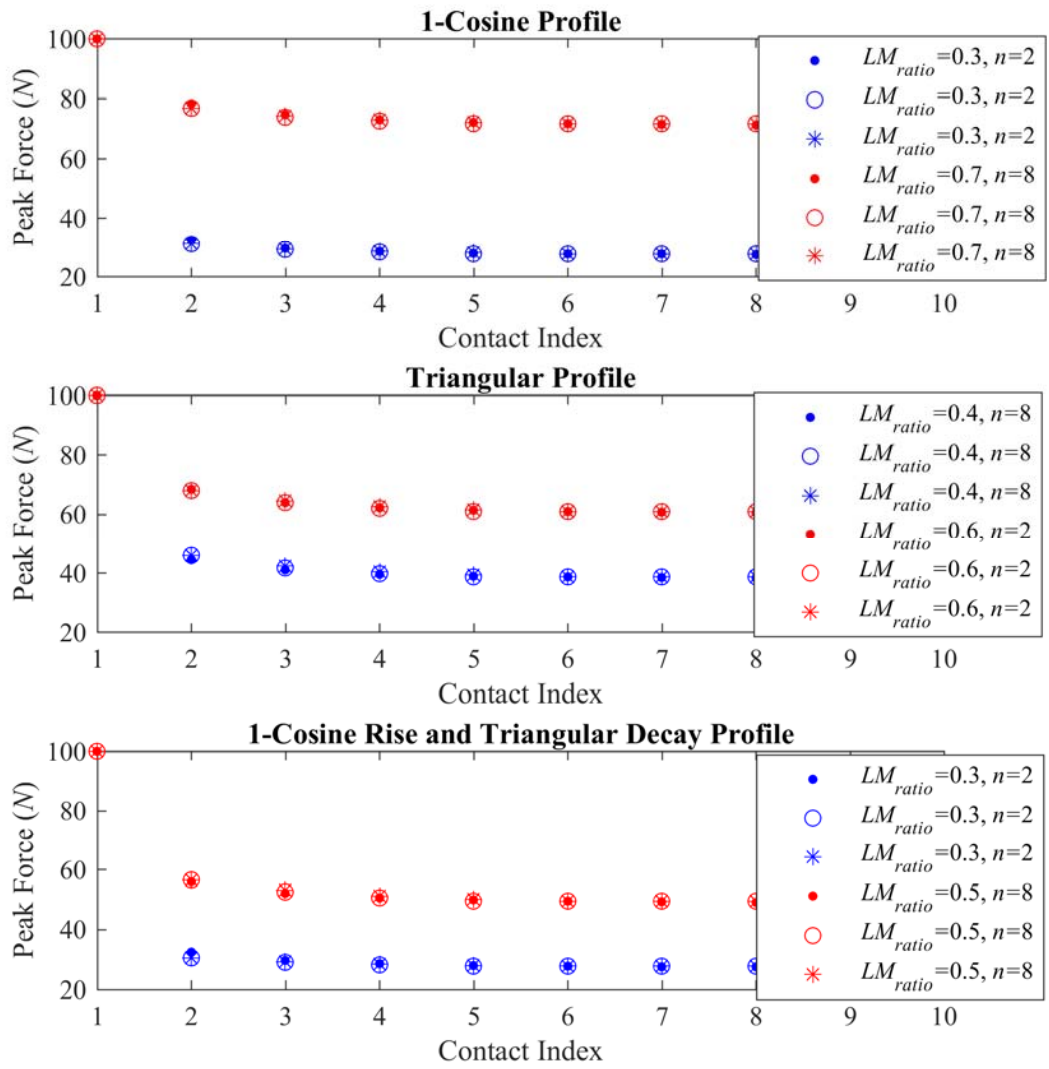

(a) $LM_{ratio} < 1$

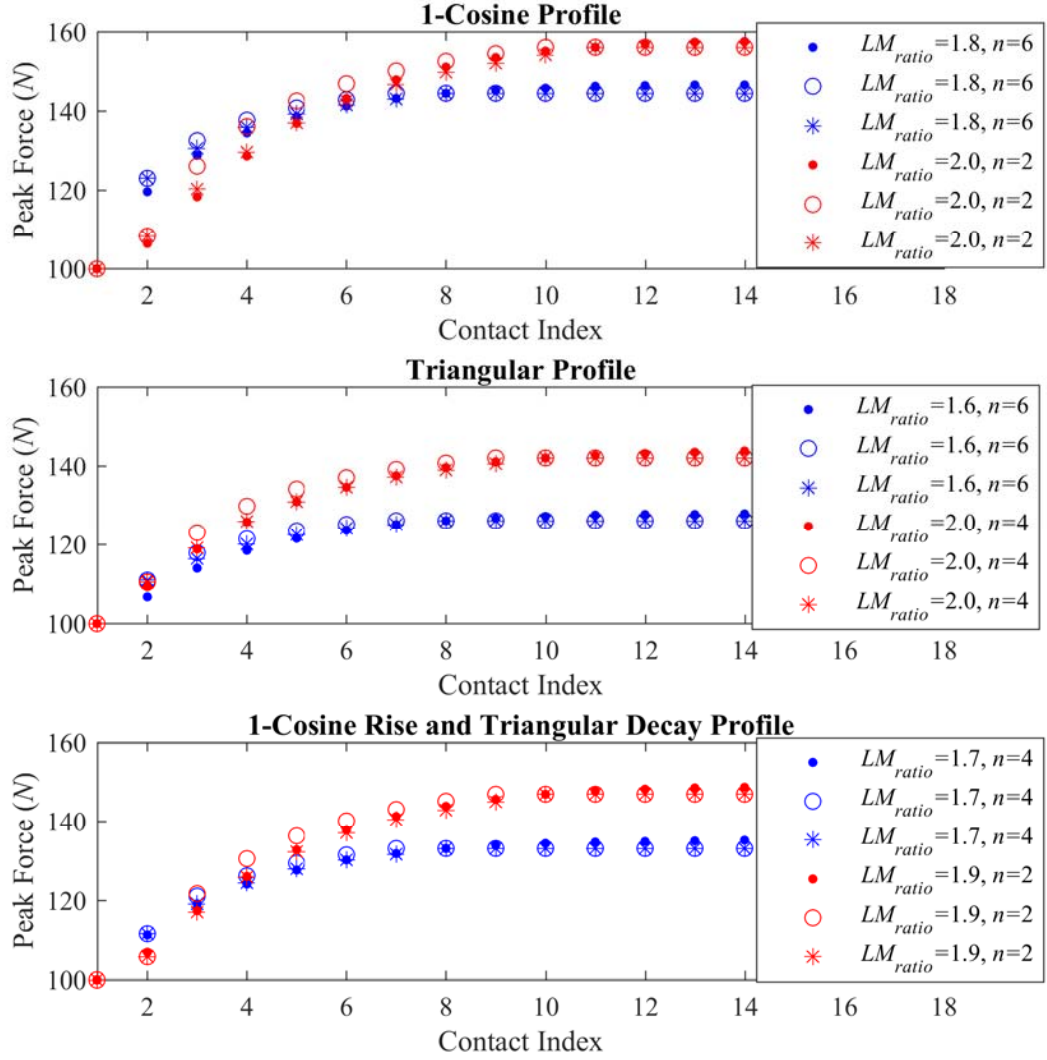

(b) $LM_{ratio} > 1$

Figure 5. Comparison between the exact and approximate peak values of the compressive contact force at each neighboring granules in the transition regime, for different linear momentum ratios, $LM_{ratio}$ and different input force profiles; here "·" denotes the exact numerical value, "o" represents the result from Equation (6), and "*" represents the result from Equation (9). System parameters: $F_0 = F_{1_{max}} = 100N$; (a) $LM_{ratio} < 1$, and (b) $LM_{ratio} > 1$.

## 6. Summary

The growth of SWs under finite rise and decay duration shocks is discussed. It is observed that the number of the granules, $p$, in the transition regime before a SW is formed depends on the type of the input force and the properties of the contacting granules. Explicit expressions are given to estimate the number $p$ of granules in the transition regime. Since, in an array of spherical contacting granules, a SW spans over 5 granules, the number $p$ will always be equal to or greater than 5. Remarkably, once the value of $p$ is known, we can also analytically estimate the peak values of the compressive contact forces in the transition regime. Numerical simulations are used to demonstrate the accuracy of the analytical expressions.

**Acknowledgement**

This research has been conducted at the Center of Excellence for Advanced Materials (CEAM) at the University of California San Diego, under DARPA grant RDECOM W91CRB-10-1-0006 to the University of California, San Diego.